\documentclass{article}
\usepackage[T1]{fontenc} 
\usepackage[utf8]{inputenc} 
\usepackage{ismir,amsmath,cite,url}
\usepackage{graphicx}

\usepackage{color}
\usepackage[ruled,noend,linesnumbered]{algorithm2e}

\usepackage[firstpage]{draftwatermark}
\definecolor{lightgray}{rgb}{0.9,0.9,0.9}
\definecolor{darkgray}{rgb}{0.4,0.4,0.4}
\SetWatermarkFontSize{12pt}
\SetWatermarkScale{1.1}
\SetWatermarkAngle{90}
\SetWatermarkHorCenter{202mm}
\SetWatermarkVerCenter{170mm}
\SetWatermarkColor{darkgray}
\SetWatermarkText{Late-Breaking / Demo Session Extended Abstract, ISMIR 2024 Conference}

\newcommand{\code}{\texttt}

\def\lbd{}	

\usepackage{lineno}
\usepackage{booktabs}
\usepackage{bm}

\title{Zero-shot Crate Digging $\rightarrow$ DJ Tool retrieval using Speech Activity, Music Structure and CLAP embeddings}

\oneauthor
 {Iroro Orife}
 {{\tt iroro@alumni.cmu.edu}}

\def\authorname{I. Orife}

\begin{document}

\maketitle
\begin{abstract}
In genres like Hip-Hop, RnB, Reggae, Dancehall and just about every Electronic/Dance/Club style, DJ tools are a special set of audio files curated to heighten the DJ's musical performance and creative mixing choices. In this work we demonstrate an approach to discovering DJ tools in personal music collections. Leveraging open-source libraries for speech/music activity, music boundary analysis and a Contrastive Language-Audio Pretraining (CLAP) model for zero-shot audio classification, we demonstrate a novel system designed to retrieve (or rediscover) compelling DJ tools for use live or in the studio.
 \end{abstract}

\section{Introduction}\label{sec:introduction}
When DJs are mixing for a live audience, or working in the studio on special edits,  remixes, re-drums, mashups or simply long-playing mixtapes, DJ tools provide a host of creative possibilities. Tools vary by genre and era, but are generally short, simplified musical phrases retrieved from existing music with the intention to reuse in a DJ performance. These musical phrases range from sound effects to acapella loops to purely instrumental passages, solo percussion or drum break to an entire verse or bridge of song. For example a DJ might trigger an acapella loop or long sound effect while mixing a transition from SongA $\rightarrow$ SongB. DJ Tools are commonly sold in online shops along with royalty-free sound libraries, sample packs of loops and beats. Most tools include key signature, beat and tempo metadata where necessary to ensure sync to the DJ project master tempo.

\subsection{Crate digging \& a short history of DJ tool}\label{sec:history}

Before the advent of online shops trading sonic tools, DJs and producers were known to spend time in record shops crate-digging, or hunting for rare, vintage, or otherwise obscure vinyl with interesting breaks, melodic hooks, drops, intros/outros, or B-side acapellas. Practise time was devoted to studying the structure of music, identifying suitable mix points, curating tools and experimenting with different creative interpolations between two mixable songs. Then while playing live, tools are triggered or looped from Sampler modules or ``Remix Decks'' connected to the DJ mixing board.

\subsection{Musical structure, speech activity and zero-shot classification}\label{sec:djtool_classes}
DJ tools naturally occur at moments in a song where there is a transition to a simpler, less-dense mix. Ergo, we leverage the music structural analysis framework (MSAF) boundary detection algorithms to supply the approximate time-offsets of structural progressions \cite{nieto2016systematic}. Next, we employ a speech and music activity detector (SMAD)\cite{Hung2022} to further refine the selection of suitable passages.

For DJ-tool classification we engage the zero-shot capabilities of a Contrastive Language-Audio Pretraining (CLAP) model \cite{elizalde2022claplearningaudioconcepts}. Given an audio segment $X^{a}$ and a list of text descriptions of different DJ tool classes \{$X^{t}_{1},...,X^{t}_{M}$\}, we use CLAP's pretrained \{audio, text\} encoders and their projection layers to compute CLAP embeddings $E^a$ and $E^t_i$. The classification logits $D_i$ can be computed as magnitude of the cosine similarity between the audio segment embedding and \textit{each} text embedding.

\begin{equation}
\begin{matrix}
    X^{a} \rightarrow AudioEncoder \rightarrow E^{a} \\ 
    \{X^{t}_{1},...,X^{t}_{M}\} \rightarrow TextEncoder \rightarrow \{E^{t}_{1},...,E^{t}_{M}\} \\ 
\end{matrix}
\end{equation}

\begin{equation}
    D_i = Similarity(E^a, E^{t}_{i}) \\ 
\end{equation}

\begin{figure}
 \centerline{
 	\includegraphics[alt={SMAD \& MSAF Analysis signals}, width=0.9\columnwidth]{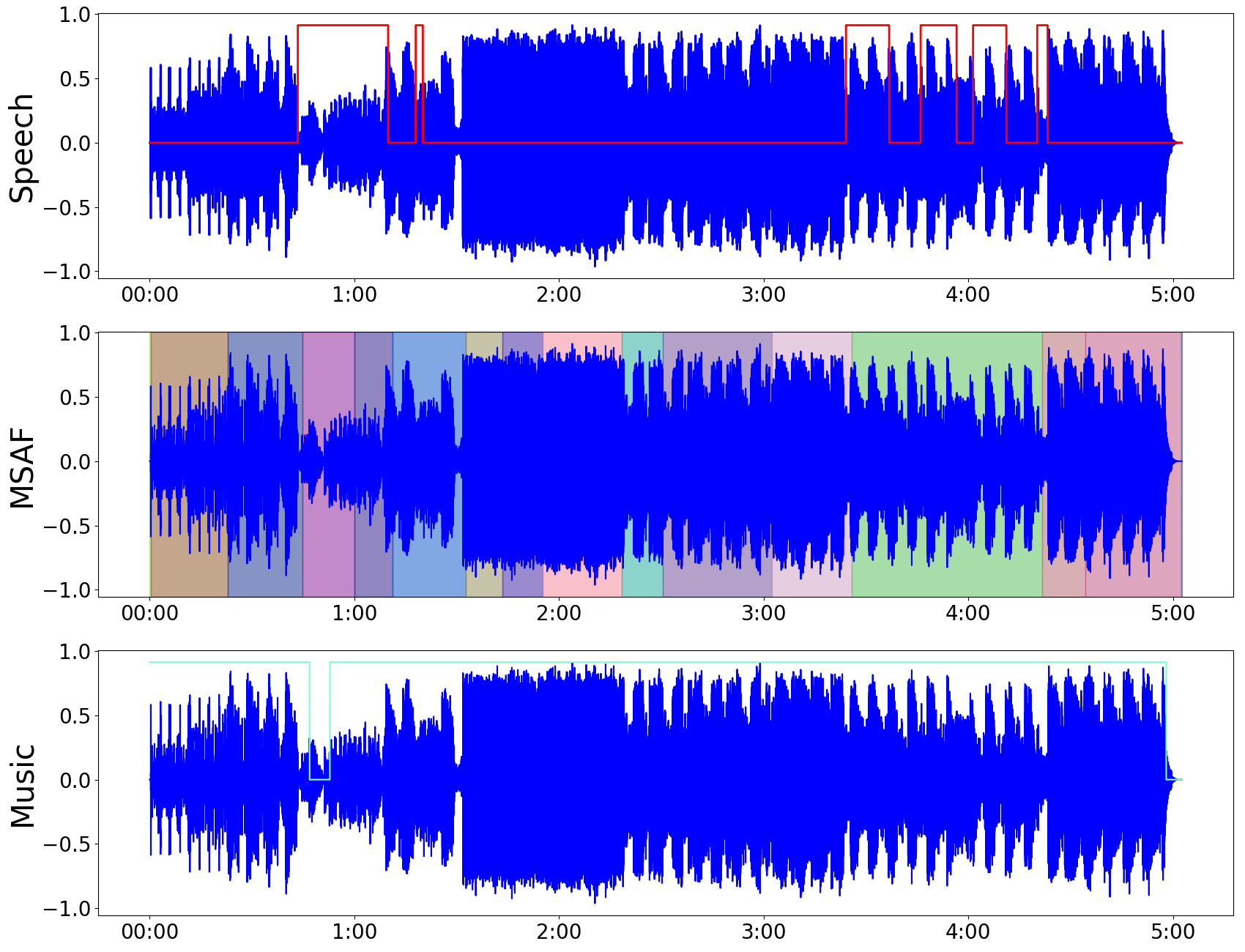}}
 \caption{A 5 minute Ragga Jungle song overlaid with detected speech and music activity, as well as music-structural boundaries}
 \label{fig:msafsmadplot}
\end{figure}

\section{The System}\label{sec:system}

The system, its libraries, data and model dependencies, are all fully open-source software (OSS) written in Python.\footnote{https://github.com/ruohoruotsi/djtool-crate-digging} Firstly, we run MSAF\footnote{https://github.com/urinieto/msaf} and SMAD\footnote{https://github.com/biboamy/TVSM-dataset/tree/master} to generate raw activities and structural boundaries, saved as CSV files. Next, these are post-processed to yield lists of time ranges and boundary times. In principle, we can just use the boundary times to segment the song, however using speech activity onset times delivers more precise slices. Figure \ref{fig:msafsmadplot} shows the relationship between the boundaries and the speech activity for one song. For zero-shot classification, we utilize the LAION version of CLAP\footnote{https://huggingface.co/laion/clap-htsat-unfused} hosted on Huggingface\cite{WuClap2023}. Finally, requiring some creativity and trial-and-error, we manually create a list of descriptive text strings for each class. A few abbreviated DJ tool descriptions are listed in Table \ref{tab:djtool_texts}.

\subsection{Algorithm}\label{subsec:algo}
Below we outline how we go about processing all files in a music library. $W^{s}, W^{m}$ are ``windows'' or time ranges specified with a \code{start\_time, end\_time, label}

\begin{algorithm}
    \caption{Zero-shot Crate Digging}\label{combo_algo}
    	Create text prompts for $M$ classes \{$X^{t}_{1},...,X^{t}_{M}$\} \\
    	 $\{E^{t}_{1},...,E^{t}_{M}\} \gets TextEncoder(\{X^{t}_{1},...,X^{t}_{M}\})$ \\
		\BlankLine

        \For{song $s_i$ in the library $S_N$}{
            $W^{s}, W^{m} \gets SMAD(s_i)$ \\ 
            $B \gets MSAF(s_i)$  \\
            Adjust boundaries $B_i$ to nearest $W^{s}$ \\
			Use $B$ to cut $s_i$ to audio files \{$X^a_1$,..., $X^a_N$\} \\
		    \BlankLine
            \For{$j := $1 to $N$}{
                $E^{a}_j \gets AudioEncoder(X^a_j)$ \\
                $\vec{D} \gets Similarity(E^a_j, \{E^{t}_{1},...,E^{t}_{M}\})$ \\
                $\vec{\bm{z}} \gets softmax(\vec{D})$ \\
                Store predicted class, the $\mathop{\arg \max}$ of $\vec{\bm{z}}$\\
            }
        }
\end{algorithm}

\subsection{Evaluation and discussion of limitations}
To evaluate the system, we generated classifications from the author's DJ library. Overall, the system performs well for vocal and percussive tool classes with ``prediction probabilities'' > $0.75$. However, shorter burstier sound effects, drops and genre-specific tools failed to be identified consistently. As expected, the system is sensitive to \textit{precise language} and we observe that adding broad or generic terms or entire genres (e.g. ``hip-hop, funk, drum and bass'') to a single class description led to markedly worse results across all classes. For best results, the classes should not overlap in their description, but can contain mixed elements. For example our best performing vocal description was ``acapella, expressively sung human vocal with background instrumental music tracks''. 

DJ tool durations are typically at the motif or phrase level, i.e. longer than a single utterance, note or beat but shorter than an entire verse or chorus. The challenge was to determine optimal segmentation for CLAP classification, while slicing the song appropriately for use by DJs. So we ran an experiment, using three 23s segments from ground-truth-ed songs \cite{MSS00401998, 91vocals2024}, tracking how the predicted class probabilities varied as we progressively reduced the segment duration to 3s.  We observe that for tools with just one class \{\textit{02\_drums.wav}, \textit{03\_vocalhook.wav}\} featuring breakbeats and purely sung vocals respectively, that predictions are stable even at short durations. Whereas for \textit{01\_vox.wav} which includes vocals,  stuttery vocal effects and eventually background music, its predictions are more sensitive to local features and variances.

\begin{table}
 \begin{center}
 \begin{tabular}{ll}
  \midrule
  \textbf{DJ Tool class} & \textbf{Example text description} \\
  \midrule
	Acapella loops & ``expressively sung vocal tracks''  \\
	Sound effects &  ``siren, riser sound effects, whoosh''\\
	Drums breaks  & ``drum beat, drum solo, breakbeat''  \\
	Melodic hooks & ``strings, solo guitar, piano melodies''  \\
	DJ Drops & ``a high energy, massive EDM drop''  \\
	Battle tracks & ``vinyl scratch loop, turnatablism''   \\
 \end{tabular}
\end{center}
 \caption{In practise, text descriptions are more tortuous}
 \label{tab:djtool_texts}
\end{table}

\begin{table}
 \begin{center}
 \begin{tabular}{llllll}
    \toprule
 	  Sliced segment & 23s & 18s & 13s & 8s & 3s \\
 	\midrule
	\textit{01\_vox.wav} & 1. & 1. & 0.99 & 0.25 & 0.01 \\
	\textit{02\_drums.wav} & 1. & 1. & 1. & 1. & 1. \\
	\textit{03\_vocalhook.wav} & 1. & 1. & 1. & 0.99 & 0.96\\
	\end{tabular}
\end{center}
 \caption{Segment predictions, varying segment duration}
 \label{tab:vox_class_ablation}
\end{table}

\section{Related Work}
There is not much existing work within the community relevant to DJs. The few studies that we found focused on Pop or EDM genres, which have much less of a crate-digging history. These works focused on the challenges of ideal sequencing of songs and inter-song transitions, with an eye toward the dream of a fully automatic DJ \cite{kim2017automatic, huang2017djnet, bittner2017automatic}.

In a recent work, DJ StructFreak (2023), the authors tackle the task of carefully choosing suitable ``mix points'' \textit{within} a song, to generate smooth, pleasing transitions \cite{kim2023dj}. Similar to our work, they employ music structural analysis and use embeddings from a pretrained model \cite{kim2023all}. 

\section{Future Work}
There is much work to be done to develop the system into a tool for non-programmer disc jockeys. Future algorithmic work includes improving the fidelity of the boundary detection algorithms and evaluating recent structural segmentation approaches\cite{kim2023all}. Finally, if there is enough interest to turn DJ tool retrieval into a proper MIR task, then we will need labeled datasets. 
\bibliography{ISMIRtemplate}

\end{document}